\documentclass[aps,prl,preprint,showpacs,showkeys,groupedaddress,amsmath,amssymb]{revtex4}

\usepackage{graphicx}
\usepackage{dcolumn}
\usepackage{bm}

\begin{document}


\title{Infra-Red Surface-Plasmon-Resonance technique for biological studies}

\author{ V. Lirtsman, M. Golosovsky,
\footnote{electronic address: golos@vms.huji.ac.il}
D. Davidov,}
\affiliation{The Racah Institute of Physics, The Hebrew University of Jerusalem, 91904 Jerusalem, Israel\\}
\date{\today} 

\begin{abstract} 
We report on a Surface-Plasmon-Resonance (SPR) technique based on Fourier -Transform - Infra - Red (FTIR) spectrometer. In contrast to the conventional surface plasmon technique, operating at a fixed wavelength and a variable angle of incidence, our setup allows the wavelength and the angle of incidence to be varied simultaneously. We explored the potential of the SPR technique in the infrared for biological studies involving aqueous solutions. Using computer simulations, we found the optimal combination of parameters (incident angle, wavelength) for performing this task. Our experiments with physiologically important glucose concentrations in water and in human plasma verified our computer simulations. Importantly, we demonstrated that the sensitivity of the SPR technique in the infrared range is not lower and in fact is even higher than that for visible light. We emphasize the advantages of infra red SPR for studying glucose and other biological molecules in living cells.
\end{abstract}
\pacs {}
\maketitle

\section{introduction}
In recent years, in the fields of biochemistry and pharmacology, there has been a rigorous and extensive search for label-free techniques capable of quantitatively monitoring biomolecules and their interaction with living cells \cite{1,2}. This can be achieved by measuring the refraction index of solutions with biomolecules. The surface plasmon resonance (SPR) technique is extremely sensitive to small variations in refractive index. SPR is usually excited in the regime of the attenuated total reflectance (ATR) when a p-polarized electromagnetic wave is incident on a metal-coated prism at a certain angle (Kretschmann's geometry) \cite{3}. The reflectivity displays a sharp minimum at certain angles/wavelengths that strongly depend on the refraction index of the dielectric medium in contact with the metal. By measuring this angle/wavelength, one can monitor in real time the concentration of a chemical substance in solution \cite{4,5,6,7} and in the cell culture as well \cite{2,8}. Most of the SPR techniques utilize glass-based optics, which limits its operation to the visible and NIR range \cite{2,4,9,10,11}. Here we examine the advantages offered by the SPR technique in the mid-infrared range, up to 12 $\mu$m. These advantages are as follows:
\begin{itemize}
\item   Fast multi-wavelength measurements. The ability to detect SPR at varying wavelengths and/or varying angles allows "tuning" the surface plasmon resonance to any desired spectral range in order to achieve the highest sensitivity. 
\item    Since many biomolecules have specific absorption bands in the infrared (so-called "fingerprints"), the tuning of SPR to this spectral range allows these biomolecules to be identified. 
\item The relatively large penetration depth of the surface plasmon into a dielectric medium (a few microns in the infrared range, which is of the order of the cell height, as compared to 0.5 $\mu$m in the visible range), is beneficial when studying cell cultures.
\end{itemize} 
The main disadvantage of the IR range when using the SPR technique is the large water absorption and lack of convenient laser sources. At first glance, this may result in a lower sensitivity with respect to biomolecules in aqueous solutions. However, we demonstrate here that by clever choice of parameters (wavelength, angle of incidence, conducting film thickness), one can achieve high enough sensitivity. To determine the sensitivity of our SPR-FTIR technique, we chose glucose in water as a test case. Briefly, we measured small and physiologically important glucose concentrations (1-50 mM) in water as well as in human plasma. We achieved a sensitivity that is comparable or even higher than that for the visible range. This may suggest that the SRR-FTIR technique is an excellent tool for studying biomolecules and living cells in aqueous solutions. 

\section{Sensitivity considerations: optimal parameters for sensitive detection of glucose in aqueous solution using SPR in the infrared}
The SPR sensitivity is limited by conductor losses and by absorption and scattering in the dielectric. In water, optical absorption in the visible range is low and the sensitivity of the SPR technique is limited by conductor loss in the metallic film. Silver is the best metal for the SPR technique in the visible range. In the infrared range, however, the situation is different. Conductor loss in the infrared range is low and the best choice of a metal for $\lambda>1.5 \mu$m is gold, although other metals, such as silver and copper should also work well. However, the sensitivity of the SPR technique in the infrared range is limited by the dielectric losses in water. Therefore, to achieve high sensitivity for a certain type of biomolecule (glucose in our present work), one should carefully choose the wavelength range where water absorption is not high enough. 
Figure 1 shows the real and imaginary parts of the refraction index $n + ik$ of pure water and dry glucose calculated based on the data of Ref. \cite{12,13}. We noted that across the wavelength range from 2 to 10 $\mu$m the water absorption is high, compared with that of glucose. Therefore, absorption spectroscopy is inefficient in measuring glucose in water. However, note the large differences in the real part of the refraction indices of water and glucose, especially in the range of 2.5-3 $\mu$m. By measuring the difference in the real part of the refraction indices of pure water and of the water-glucose solution, we can measure the glucose concentration. The sensitivity of the SPR technique to small variations in the refractive index depends is a complicated way on conductor thickness $d$, wavelength $\lambda$, incident angle $\Theta$, and the average refraction index of solution, $n_{solution}$  \cite{9,10,14,15} so choosing the most sensitive configuration represents a serious task. In the following paragraphs we discuss how to find the optimal combination of $\lambda$ and $\Theta$ in order to achieve the highest sensitivity to glucose in water. A similar strategy can be applied to biomolecules other than glucose.
In our experiments the surface plasmon is excited using Kretchmann's geometry based on a ZnS prism coated with a 12 nm thick gold film. To optimize the sensitivity of this setup, we studied numerically how the reflectivity of the ZnS/Au/water interface, $R_{water}$, depends on the wavelength and incident angle. The simulations were based on the Fresnel reflectivity formulae \cite{3}. Figure 2 shows computed reflectivity versus wavelength and incident angle. Note the resemblance to Fig. 1, where the vertical scale in Fig.1 (refractive index) corresponds to the vertical scale in Fig. 2 (incident angle). Indeed, the sharp black areas corresponding to the surface plasmon resonance roughly follow the spectrum of the refractive index of water in Fig.1. Note also that the surface plasmon disappears at frequencies corresponding to strong water absorption bands (1600 cm$^{-1}$ and 3400 cm$^{-1}$, see thick arrows in Fig.3).
To estimate the optical reflectivity of the water-glucose solution, we need to know its complex refractive index over a wavelength range of 2-12 $\mu$m. Our estimate is based on effective-medium approximation \cite{16} 
\begin{equation}
\epsilon_{solution}=\epsilon_{water}\left(1+3c_{glucose}\frac{\epsilon_{glucose}
-\epsilon_{water}}{\epsilon_{glucose}+2\epsilon_{water}}\right)
\label{eq1}
\end{equation}

In the limit of small concentrations, $c <15\%$ (that corresponds to 83.3 mM), Eq.\ref{eq1} yields a linear dependence on concentration. In particular, at $\lambda=2.5\mu$m the refraction index is $n=1.25+ 4.2·10^{-5}c$, whereas at $\lambda=0.586 \mu$m, it is $n=1.33+ 2.57·10^{-5}c$ (here, $c$ is in mM). This linear dependence for the visible range has been verified experimentally in several works \cite{5,6,17}; hence, we believe that Eq.\ref{eq1} is valid for the infrared range as well. 

In the next step, we calculated the reflectivity for a similar system, but we replaced the pure water with a 0.3$\%$ D-glucose solution. We analyzed the difference, $\Delta R= R_{water}-R_{solution}$, and found that it is most pronounced at the wavelength-angle range, which corresponds to the surface plasmon resonances (black areas in Fig.2). The highest $\Delta R$ was achieved for $\Theta=34^{0}$ and $\lambda\sim$ 3700-5000 cm$^{-1}$. In principle, $\Delta R$ can be further increased by the proper choice of Au film thickness \cite{9}. 
Next, we estimated the SPR sensitivity to refractive index variation. Note that under conditions of surface plasmon resonance, the dependence of the reflectivity on the refraction index is a nonlinear function that can be linearized only in a very small range of refraction index variations. To estimate the sensitivity, we calculated the reflectivity of a hypothetical solution that has the same loss tangent as water but whose refraction index is slightly different. Figure 3 shows that $S=dR/dn$ (at a fixed angle where $\Theta=34^{0}$) strongly depends on the average refraction index. The maximum sensitivity is $S_{max}= $75 RIU$^{-1}$; for pure water this is achieved at $\lambda=2.49 \mu$m. When the average refraction index of the solution deviates from that of the water by more than $0.5\%$, the sensitivity decreases by a factor of two, which defines the dynamic range of the SPR technique. In the context of glucose in water, this means that if we wish to measure the glucose concentration in water by measuring infrared reflectivity under surface plasmon resonance and using the linear relation $R=R_0+S_{max}cdn/dc$, we are limited  to glucose concentrations less than  $c_{glucose}< $50 mM. If we wish to measure higher glucose concentrations or to measure glucose in media with a slightly different average refraction index (for example, human plasma), we must use a different angle of incidence or a different wavelength ($\lambda=$ 2.48 $\mu$m in our case).

\section{Experimental}
Figure 4 schematically shows our experimental setup. A tungsten lamp of the Bruker Equinox 55 FTIR spectrometer, equipped with the KBr beam splitter, served as our light source. Briefly, the infrared beam is emitted from the external port of the spectrometer and passes through a collimator, consisting of a 1-mm diameter pinhole mounted between two gold-coated off-axis parabolic mirrors with a focal length of 76.2 mm and 25.4 mm, correspondingly. The diameter of the collimated beam is 4 mm. This beam passes through the grid polarizer (Specac, Ltd.) and is reflected from the right-angled gold-coated ZnS prism (ISP Optics, Inc.) mounted on a rotating table. Another parabolic mirror focuses the reflected beam onto the liquid-nitrogen-cooled MCT (HgCdTe) detector that is mounted on a separate rotating table. A temperature-stabilized flow cell with a volume of 0.5 mL is in contact with the gold coating. The gold film thickness is $d_{Au}$=12-18 nm.
To operate this setup, first we choose the optimal conditions (incident angle, wavelength, and gold film thickness) to achieve maximum sensitivity for a given task. Then, we fill the flow cell with a solution that includes the biomolecules (glucose), and we measure the reflectivity spectrum first for the $s$- and then for the $p$-polarization. The FTIR collects the data with the 4 cm$^{-1}$ resolution and averages the data over 16 scans. The reflectivity of the $s$-polarized beam is used as a background, in such a way that the output of our experiment is the ratio of reflectivities for $p$- and $s$-polarizations. We perform the series of measurements for ascending glucose concentrations, followed by a measurement for pure solvent, and then analyze the difference, $\Delta R=R_{solution}-R_{solvent}$.

\section{Experimental results and data processing}

Figure 5 compares reflectivity spectra for the double-distilled water and for the 55.5 mM D-glucose solution in water. The surface plasmon resonance for the pure water manifests itself as a deep minimum at 4350 cm$^{-1}$. (The shallow minimum at 5173 cm$^{-1}$ corresponds to a weak absorption band of water.). For the glucose solution, the reflectivity minimum is slightly shifted. Generally, to quantify the changes in the SPR spectra, one considers the SPR shift \cite{6,14,18}. Although this is well justified for those measurements at constant wavelengths and variable angles where the SPR dip is rather sharp; such a procedure is less justified for those measurements with variable wavelengths, where SPR is wide and the variation in the SPR width is appreciable. In this case, the whole SPR reflectivity curve should be analyzed \cite{9,10,19}. Therefore, we analyzed the differential spectra $\Delta R=R_{solution}-R_{solvent}$  (Fig. 6).
The differential spectra for small glucose concentrations are very similar and in fact can be scaled (see inset to Fig.6). The scaled spectrum is a "fingerprint" of glucose in water whereas the scaling factor is a measure of the glucose concentration. We calculate this scaling factor at two points in the differential spectra to achieve better accuracy in glucose concentrations. One of them is the minimum itself and another point is at 4900 cm$^{-1}$. The corresponding data are plotted in Fig. 7. The differential reflectivity is linearly proportional to the glucose concentration with very high exactness. We performed similar measurements at several angles of incidence ($32^{0}-36^{0}$) and found a linear dependence on concentration as well, whereas the highest sensitivity was achieved at $\Theta=34^{0}$. From the slope of the linear dependence in Fig. 7, we found sensitivity, $S_{exp}=|dR/dc|=3\cdot 10^{-5}$ mM$^{-1}$. This is in good agreement with the computed sensitivity $S_{sim}= 3.2\cdot 10^{-5}$ mM$^{-1}$.
We performed similar measurements with the D-glucose solution in human plasma. Here SPR is 200 cm$^{-1}$ red-shifted with respect to SPR in pure water due to the presence of salts, proteins, and other solutes (not shown here) that affect the average refraction index of solution. Therefore, the optimal wavelength in this case is $\lambda=$=4130 cm$^{-1}$. Clearly, the differential reflectivity linearly depends on the D-glucose concentration (Fig. 7, inset) as well. The slope of the curve, i.e., the sensitivity, is almost the same as for glucose in water.

\section{Discussion}
We demonstrated here that FTIR-SPR can be successfully used to measure physiologically important glucose concentrations in water and in human plasma. Next, we compare the sensitivity of our technique to similar techniques that operate in the visible range. The minimal glucose concentration that we were able to measure is 0.8 mM. This corresponds to a resolution of $3\cdot 10^{-7}$ RIU at $\lambda==2.5 \mu$m. This is better than the resolution reported by other groups for aqueous solutions and SPR techniques operating in the visible and NIR range: $8.6\cdot 10^{-6}$ RIU at $\lambda=$0.9-1.1 $\mu$m; 6 and $2.8\cdot 10^{-5}$ RIU for the visible range $\lambda=$0.5-0.7 $\mu$m \cite{5}. K. Johansen et al. \cite{10}- have shown that resolution $dR/dn$ increases with wavelength and can achieve specified values in the IR range. The ultimate resolution of SPR in air is $5\cdot 10^{-5}$ at $\lambda=$0.6 $\mu$m and $1\cdot 10^{-5}$ at $\lambda=$0.85 $\mu$m \cite{15}.
We can also compare our FTIR-SPR technique to other label-free methods measuring glucose concentration in water. In particular,  optical low-coherence reflectometry \cite{17} measures 2-10 mM glucose concentration with ± 1 mM accuracy at  $\lambda=$1.3 $\mu$m; the NIR surface-enhanced Raman spectroscopy  measures glucose concentration  in the 0.5-44 mM range. \cite{20} 

Finally, we will discuss how our FTIR SPR technique may be applied for studying living cells and their interaction with biomolecules, including glucose. We wish to show that FTIR-SPR is better suited for this task than SPR in the visible range. Indeed, the cells can be grown directly on the gold surface and by measuring the SPR reflectivity from the ZnS/Au/cell culture+solution interface, one can study the interaction of cells in a physiological solution with various biomolecules dissolved there. To achieve this goal, the penetration depth of the surface plasmon into dielectric media should be comparable to the cell height. Figure 9 shows the penetration depth of the surface plasmon \cite{3} excited at the gold-water interface. For the visible range, the SPR penetration length into the water, $\lambda\sim$ 0.25 $\mu$m, is much smaller than the cell height $h=3-12\mu$m. Therefore, the surface plasmon in the visible range penetrates the cell membrane but not the cell body. In contrast, for $\lambda=$2.5 $\mu$m, the penetration length is $\delta =4\mu$m whereas for $\lambda=$4.5 $\mu$m, it is even higher, $\delta =$12 $\mu$m. This penetration depth is indeed comparable to the cell height. Figure 9 shows that for $\lambda>$4.6 $\mu$m SPR penetrates too deep, thereby sensing largely the extracellular environment rather than the cells themselves, whereas the spectral region 2.6 - 3.1 $\mu$m is characterized by enhanced water absorption and therefore is also excluded. The spectral ranges of 0.5-2.6 $\mu$m and 3.1-4.6 $\mu$m represent a compromise between water absorption, SPR sensitivity, and penetration depth, as compared to cell height. These ranges are optimal for studying cell cultures grown on gold.

\section{Conclusions}
We demonstrated the effectiveness of the FTIR-based surface plasmon-resonance technique operating in the mid-IR range. We verified the sensitivity of this technique by monitoring small but physiologically important glucose concentrations in water. The system has a high resolution of $3\cdot10^{-7}$ RIU, which corresponds to a measurement error of 0.1 mM of glucose in water. The smallest glucose concentration that we have been able to measure is 0.8 mM. Our results suggest that the FTIR-based surface plasmon-resonance technique operating in the mid-IR range is an effective optical system for measuring glucose, and this strengthens the possibility of developing a non-invasive blood glucose device based on the SPR effect in the mid-IR range. Clearly, our technique can monitor other biomolecules as well.

\begin{acknowledgments}
This work was supported by the Israel Science Foundation (grant No. 1337/05) and by the Johnson $\&$ Johnson, Inc. We are grateful to B. Aroeti for useful discussions and to S. Shimron and V. Olshansky for their help in the experiments.
\end{acknowledgments}
\pagebreak 

\pagebreak
\begin{figure}[ht]
\includegraphics[width=0.8\textwidth]{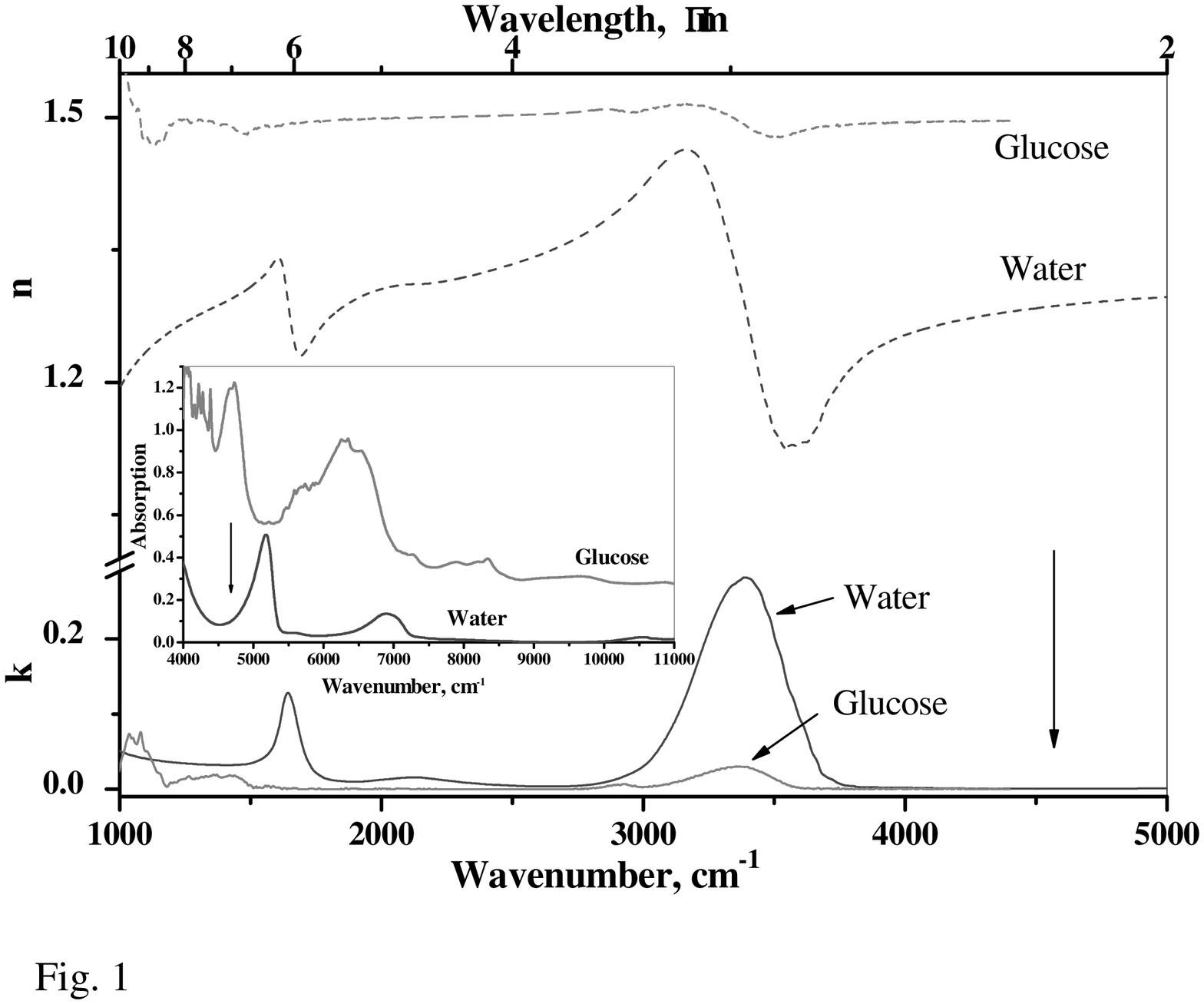}
\caption{Real $n$ and imaginary $k$) parts of the refraction index of water and of the dry D-glucose. (Based on data of Refs. \cite{12,13}). Inset shows FTIR absorption spectra of water (our measurements) and dry D-glucose (Bruker, Inc. database). By arrows we show our working place (SPR minimum) that corresponds to the glucose absorption peak (see inset) and the maximum difference between the refractive index n for water and glucose.}
\label{fig:1}
\end{figure}

\begin{figure}[ht]
\includegraphics[width=0.8\textwidth]{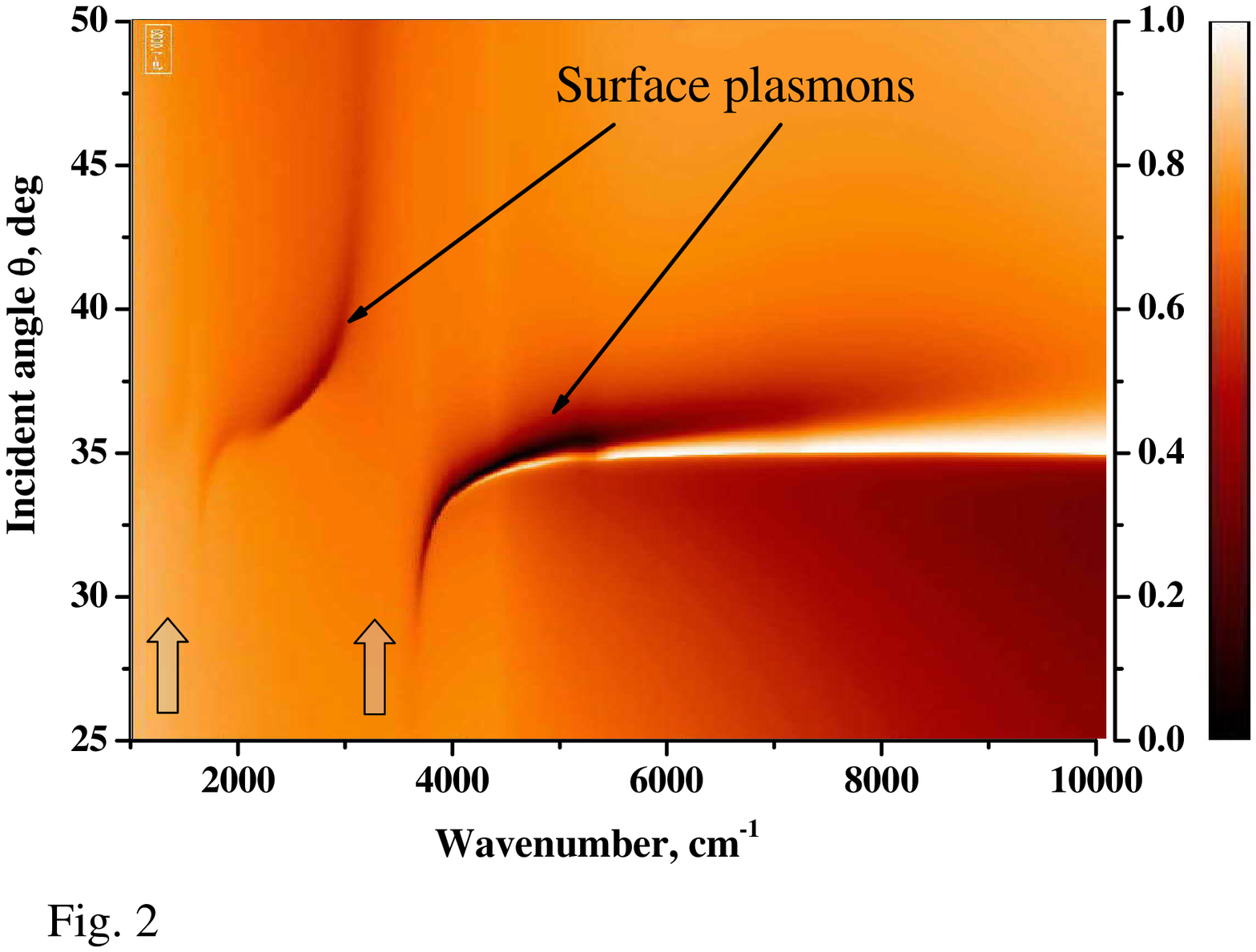}
\caption{Computer simulation based on the Fresnel reflectivity formulae of the reflectivity from the ZnS/Au/water interface versus wavelength and angle of incidence, $\Theta$ (this is the angle between the beam direction in the ZnS and the normal to the ZnS-water interface). The narrow dark areas correspond to the surface plasmon resonance. The thick arrows indicate water absorption peaks. The Au film thickness is 12 nm. }
\label{fig:2}
\end{figure}

\begin{figure}[ht]
\includegraphics[width=0.8\textwidth]{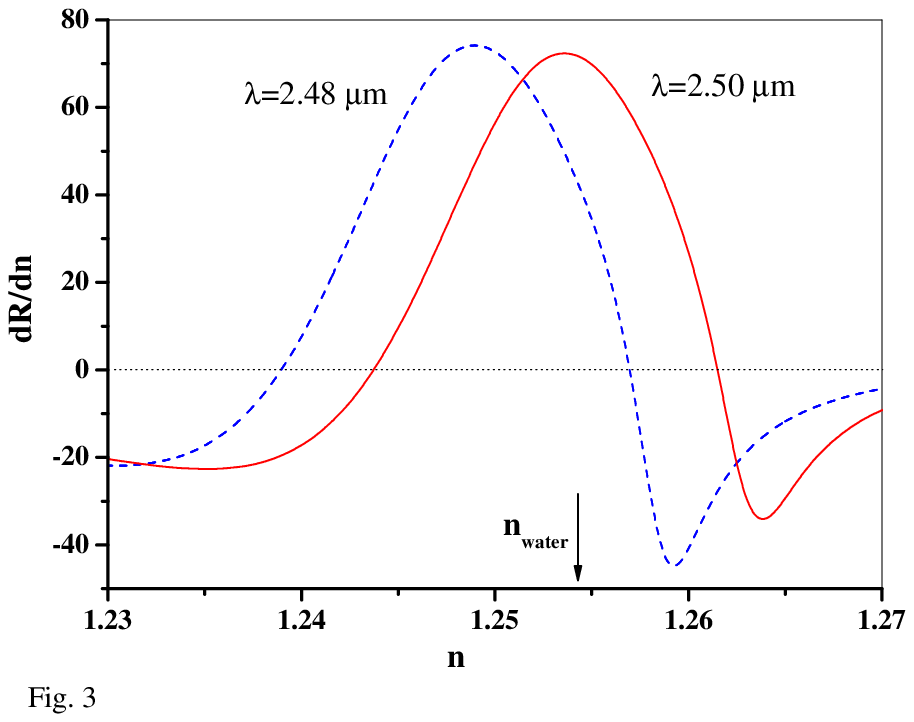}
\caption{Calculated sensitivity of the SPR technique to refraction index variations (based on Fig.2). $\Theta= 34^0$. The continuous line stands for $\lambda=$2.50 $\mu$m; the dashed line stands for $\lambda=$2.48 $\mu$m }
\label{fig:3}
\end{figure}

\begin{figure}[ht]
\includegraphics[width=0.8\textwidth]{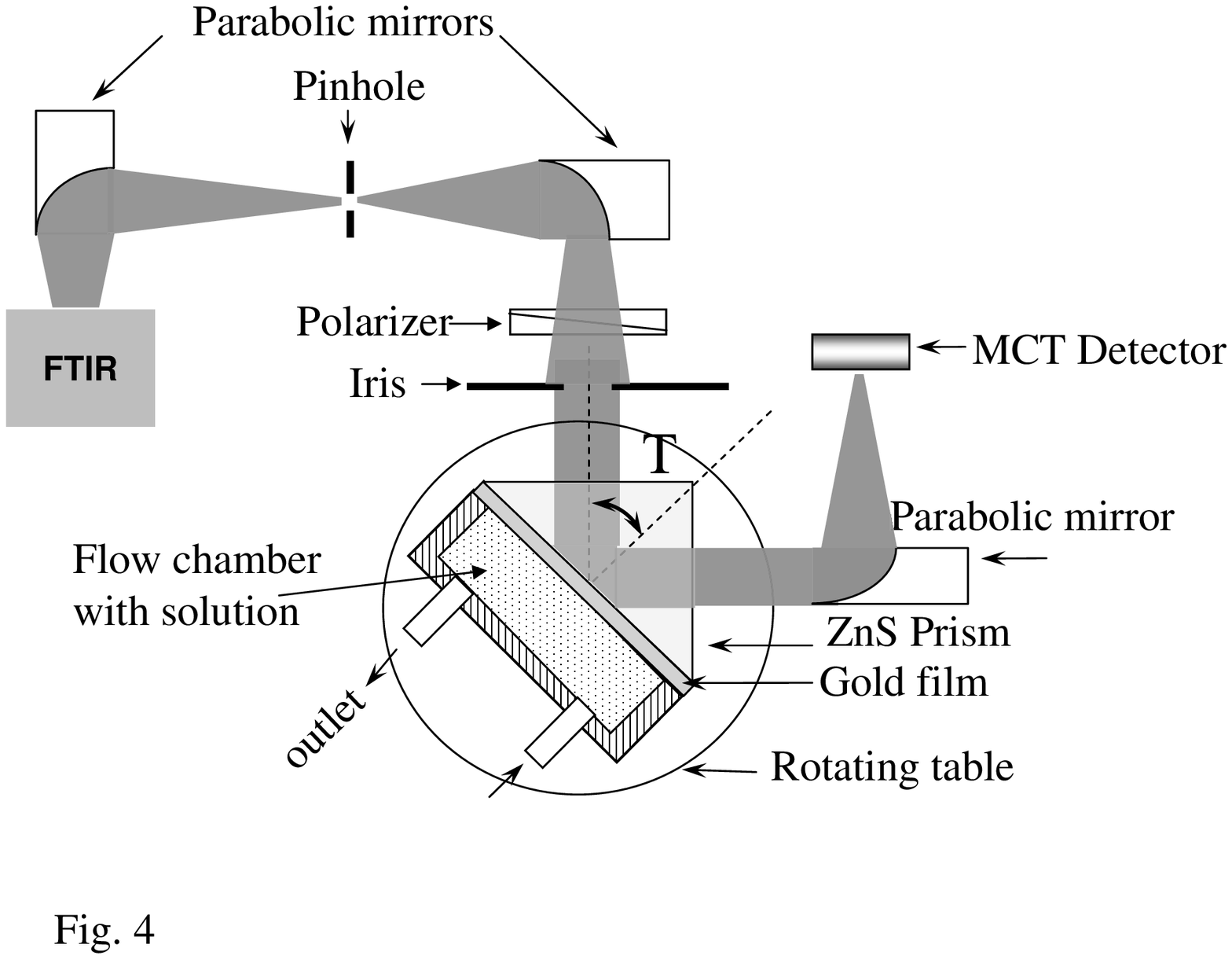}
\caption{FTIR-based Surface Plasmon Resonance technique. We measured the reflectivity spectrum from the ZnS/Au/solution interface at different angles using a FTIR source and MCT detector. The angular and wavelength ranges correspond to the excitation of the surface plasmon resonance. }
\label{fig:4}
\end{figure}

\begin{figure}[ht]
\includegraphics[width=0.8\textwidth]{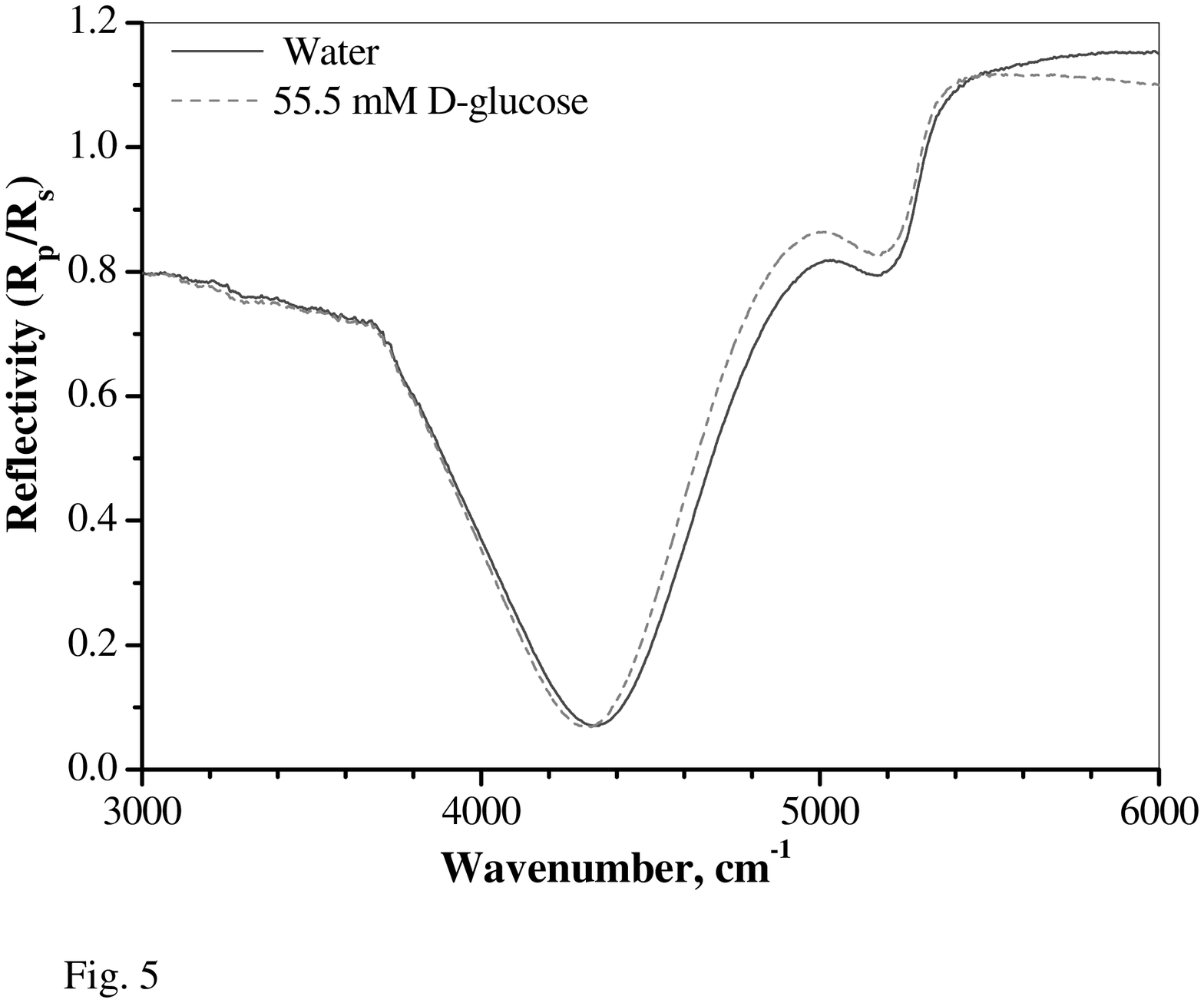}
\caption{Measured reflectivity of the water and water-glucose solution. The minimum at 4330 cm$^{-1}$ corresponds to the surface plasmon resonance. Upon addition of the glucose, the peak shifts and broadens.}
\label{fig:5}
\end{figure}

\begin{figure}[ht]
\includegraphics[width=0.8\textwidth]{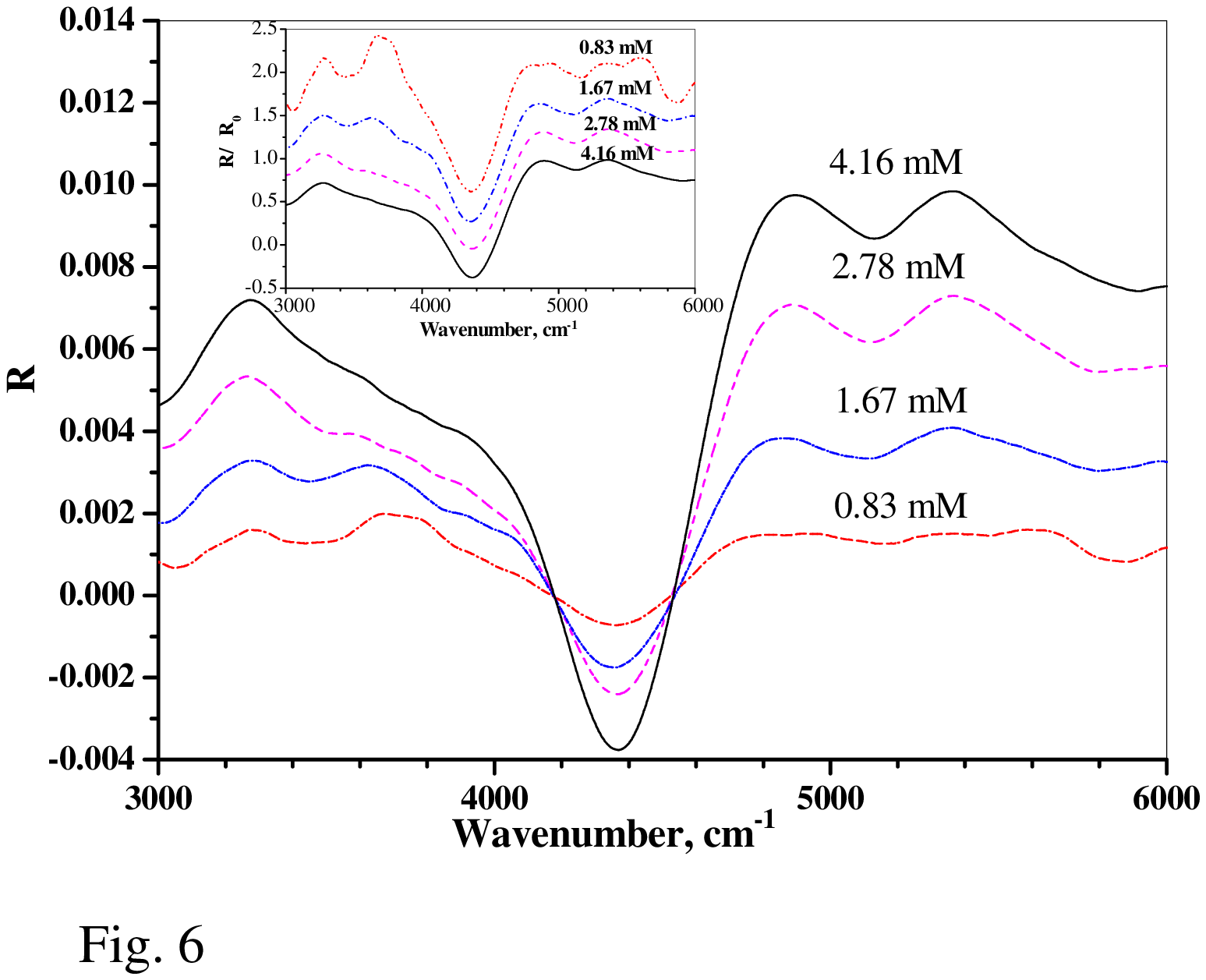}
\caption{Differential reflectivity, $\Delta R=R_{water}-R_{water+glucose}$ under the surface plasmon-resonance regime. The numbers at each curve indicate the D-glucose concentration in mM. When the glucose concentration is varied, the shape of the spectra undergoes only a very slight change.}
\label{fig:6}
\end{figure}

\begin{figure}[ht]
\includegraphics[width=0.8\textwidth]{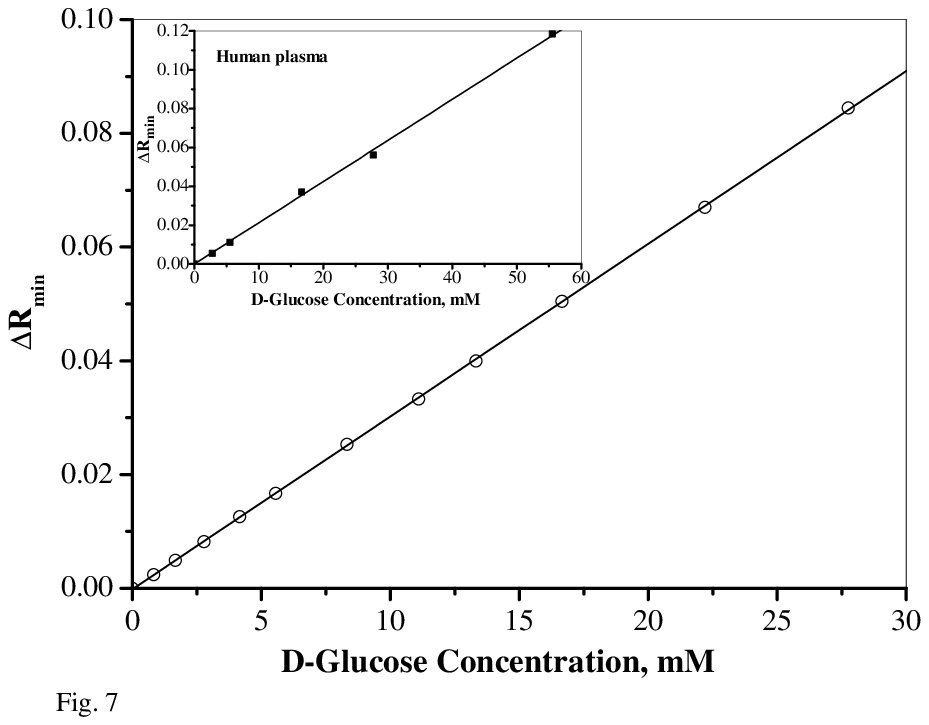}
\caption{Differential reflectivity at $\lambda$=4330 cm$^{-1}$ versus glucose concentration in water. Note the linear dependence. Inset shows the corresponding results for glucose in human plasma. Note that at the same angle the SPR minimum of human plasma is 200 cm$^{-1}$ red-shifted with respect to that of the water because of the numerous solutes that increase the average refraction index of solution.}
\label{fig:7}
\end{figure}

\begin{figure}[ht]
\includegraphics[width=0.8\textwidth]{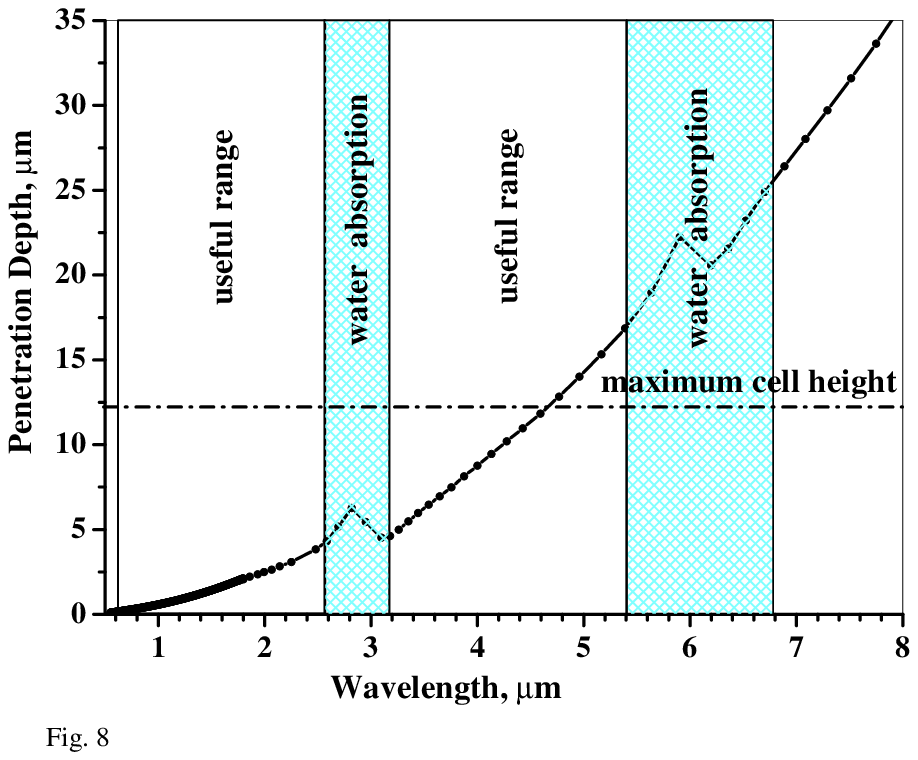}
\caption{Schematic drawing that illustrates the choice of spectral regions for the SPR-based monitoring of processes in a single layer of cells grown on a gold surface. The requirements are as follows: (i) Surface plasmon penetration depth into biomedium (solid line) is on the order of the cell height (3-12 $\mu$m); (ii) water absorption bands are excluded. This leaves the region of 0.5-2.6 $\mu$m and 3.1-4.6 $\mu$m.}
\label{fig:8}
\end{figure}

\begin{thebibliography}{}
\bibitem{1}   R. Ince and R. Narayanaswamy, Anal. Chim. Acta 569, 1 (2006).
\bibitem{2}   R. Ziblat, V. Lirtsman, D. Davidov, and B. Aroeti, Biophys. J. 91, 776 (2006).
\bibitem{3}   H. Raether, Springer Tracts in Modern Physics 111, 1 (1988).
\bibitem{4}   A. G. Frutos, S. C. Weibel, and R. M. Corn, Anal. Chem. 71, 3935 (1999).
\bibitem{5}   K. S. Johnston, M. Mar, and S. S. Yee, Sens. Actuators B 54, 57 (1999).
\bibitem{6}   W. W. Lam, L. H. Chu, C. L. Wong, and Y. T. Zhang, Sens. Actuators B 105, 138 (2005).
\bibitem{7}   T. Neumann, M. L. Johansson, D. Kambhampati, and W. Knoll, Adv. Funct. Mat. 12, 575 (2002).
\bibitem{8}   Y. Yanase, H. Suzuki, T. Tsutsui, T. Hiragun, Y. Kameyoshi, and M. Hide, Biosensors $\&$ Bioelectronics 22, 1081 (2007).
\bibitem{9}   A. Ikehata, T. Roh, and Y. Ozaki, Anal. Chem. 76, 6461 (2004).
\bibitem{10}  K. Johansen, H. Arwin, I. Lundstrom, and B. Liedberg, Rev. Sci. Instr. 71, 3530 (2000).
\bibitem{11}  V. Lirtsman, R. Ziblat, M. Golosovsky, D. Davidov, R. Pogreb, V. Sacks-Granek, and J. Rishpon, J. Appl. Phys. 98, 93506 (2005).
\bibitem{12}  M. Jetzki and R. Signorell, J. Chem. Phys. 117, 8063 (2002).
\bibitem{13}  E. D. Palik, Handbook of Optical Constants of Solids II , (Academic Press, Inc., 1991).
\bibitem{14}  J. Homola, I. Koudela, and S. S. Yee, Sens. Actuators B 54, 16 (1999).
\bibitem{15}  J. Homola, S. S. Yee, and G. Gauglitz, Sens. Actuators B, 54, 3 (1999).
\bibitem{16}  L. D. Landau and, E. M. Lifshitz Electrodynamics of Continuous Media (Butterworth-Heinenann, Oxford, 2002).
\bibitem{17}  K. V. Larin, T. Akkin, R. O. Esenaliev, M. Motamedi, and T. E. Milner, Appl. Opt. 43, 3408 (2004).
\bibitem{18}  S. Patskovsky, A. V. Kabashin, M. Meunier, and J. H. T. Luong, J. Opt. Soc. Am. A 20, 1644 (2003).
\bibitem{19}  K. Kurihara, K. Nakamura, and K. Suzuki, Sens. Actuators B. 86, 49 (2002).
\bibitem{20}  D. A. Stuart, C. R. Yonzon, X. Y. Zhang, O. Lyandres, N. C. Shah, M. R. Glucksberg, J. T. Walsh, and R. P. Van Duyne, Anal. Chem. 77, 4013 (2005).

\end{thebibliography}
\end{document}